\documentclass[aps,showpacs,amsmath,amssymb,twocolumn,final]{revtex4-1}

\usepackage{graphicx, epstopdf}

\usepackage{dcolumn}
\usepackage{bm}
\usepackage{ulem}
\usepackage{soul}

\usepackage{dcolumn}
\usepackage{bm}

\begin{document}
	
	
	\title{ Microwave response and electrical transport studies of disordered s wave superconductor: NbN thin films}
	
	\author{D. Hazra, S. Jebari, R. Albert,  F. Blanchet, A. Grimm, C. Chapelier and M. Hofheinz}
	
	\affiliation{ Univ.\ Grenoble Alpes, CEA, INAC-PHELIQS, 38000 Grenoble, France }

\begin{abstract}
It is now well known that the properties of the disordered s wave superconductors can deviate significantly from the prediction of standard Bardeen-Cooper-Schrieffer (BCS) theory. By measuring the temperature dependence of the resonance frequency, $f_0$, of microwave resonators made from disordered NbTiN and TiN thin films, at low temperatures, below half of the superconducting critical temperature, $T_c$,  Driessen et al. \cite{driessen2012PRL} demonstrated that $f_0$ vanishes faster than predicted from the BCS theory. Here, we report on the temperature dependence of $f_0$ of microwave resonators made from disordered NbN thin films at higher temperatures--- typically from 0.4 to 0.8$T_c$. In this temperature range, we demonstrate that $f_0$ vanishes slower than predicted from the BCS theory. We discuss the possible role of electronic inhomogeneity and possibility of appearing a pseudogap-like feature. We also discuss the possibility of a faster downturn of the superfluid density, $n_s$, near $T_c$, resembling a Berezinski--Kosterlitz--Thoules (BKT) type transition.

\end{abstract}

\maketitle



\section{Introduction}

It had been a long-standing paradigm that the superconducting properties are unaffected by disorder as long as the time reversal symmetry is intact \cite{anderson1959theory}. Consequently, it is expected that $T_c$ of an s-wave superconductor would be related to its energy gap, $\Delta$, by BCS relation, irrespective of its normal state resistivity. But in last few decades, theory and numerical simulations, followed by several experiments, have demonstrated that even for a moderate level of disorder there can be a significant departure from standard BCS like behaviour \cite{finkel1994physica, kowal1994disorder,  goldman1998, ghosal1998role, gantmakher1996strong, ghosal2001inhomogeneous, baturina2007prl, sambandamurthy2004superconductivity, sambandamurthy2005experimental, vinokur2008superinsulator, ghosh2013amplitude}.

Followed by these early developments, in recent times, there has been again a resurgence to study the properties of disordered superconductors and already a large wealth of physics has been unearthed --- which are not conceivable by conventional BCS theory --- like, superconductor-insulator transition \cite{goldman1998, baturina2007prl}, magnetic flux quantization in disordered driven insulating film \cite{stewart2007superconducting}, finite superfluid stiffness above $T_c$ \cite{crane2007survival}, spatially inhomogeneous superconductivity \cite{sacepe2008prl, kamlapure2013emergence, noat2013unconventional}, jump in superfluid density near $T_c$ \cite{kamlapure2010APL}, strong phase fluctuation \cite{mondal2011PRL} and localization of the preformed Cooper pairs \cite{sacepe2011natphysics}, to mention only few.

Most of these studies, however, focus on scanning tunnelling spectroscopy (STS) or magnetoresistance measurements. In comparison to that, probing disordered superconductors using microwave signal has been a few (see e.g, \cite{crane2007survival, driessen2012PRL, coumou2013PRB, mondal2013SCR, liu2011dynamical}). Moreover, recently,  the microwave resonators made of disordered superconductors have been explored to perform circuit quantum electrodynamics (circuit-QED) experiments at a high magnetic field for quantum computation and also as high impedance resonators \cite{samkharadze2016high, luthi2017evolution, landig2017coherent, maleeva2018circuit,samkharadze2018strong}--- thanks to their high kinetic inductance. These, in turn, demand to probe the disordered superconductors using the microwave as their detailed properties might impact circuit-QED experiment's outcome.  

Here, we report on the microwave response of disordered s wave superconductors: NbN thin films. For that, microwave resonators are fabricated and their resonance frequencies, $f_0$, are measured as a function of temperature. With increasing disorder, we observe a systematic departure from BCS behaviour: $f_0$ extrapolates to zero, in the BCS framework, at a temperature $T_c$ which is always higher than $T_{c1}$ where resistivity goes to zero. Moreover, the difference, $T_{c} - T_{c1}$, widens with disorder. We discuss whether the observed behaviour can be due to the electronic inhomogeneity or the presence of a pseudogap-like feature. We also analyze if a sudden downturn of the superfluid density, $n_s$, near $T_c$, resembling a BKT type transition, is behind the observed microwave response of the samples.  


\section{Experiments, results and analysis}

Three NbN thin-films are deposited by d.c magnetron sputtering of an Nb target, in the atmosphere of nitrogen and argon gas mixture, onto oxide-coated Si substrate at room temperature. Prior to NbN deposition, the substrates are cleaned by back-sputtering and a 20 nm  MgO buffer layer is added to improve the film quality. Microwave resonators, transmission lines and Hall bars are fabricated using standard optical lithography and reactive ion etching technique. The detailed fabrication processes are reported elsewhere \cite{grimm2015josephson,grimm2017self}. The electrical transport and microwave transmission ($S_{21}$) measurements are performed down to liquid helium temperature.

\begin{figure}\centerline{\includegraphics[width=7cm,angle=0]{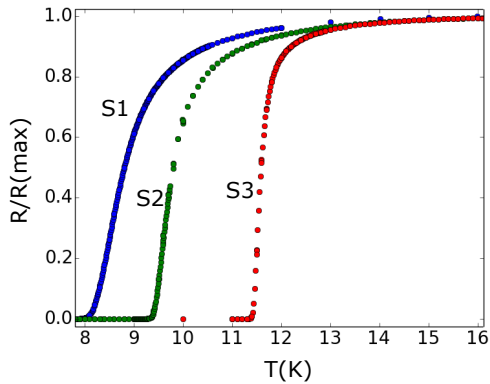}}
\caption {Resistance, R, as a function of temperature, T,  near superconducting transition for three NbN thin films we have studied. The resistances are normalized by the maximum resistance of the respective film. The transition temperatures, $T_{c1}$s, where R become zero, are listed in Table-\ref{tab1}.}
\label{fig:ResVsTemp}
\end{figure}

\begin{table*}
\caption{\label{tab1} Summary of the various parameters. See the main text for details.}
\begin{tabular}{|c|c|c|c|c|c|c|}
\hline
Samples & $R_{\Box}$ & d & $k_{F} \ell$ &  $T_{c1}$ & $T_{c}$ & $T_{c} - T_{c1}$  \\
        & ($\Omega$)& (nm) &  &  (K) & (K) & (K)   \\
S1   &1100 & $\sim$  3.5  & $\sim $1.7 & 7.85  & 9.37 &1.52  \\
S2   &540 & $\sim$  5.0 & $\sim$ 2.5 & 9.19  &10.20  & 1.01  \\
S3   &160 & 10.0 & 4.1 & 11.35  & 12.02 &0.67  \\

\hline
\end{tabular}
\end{table*}

In Fig.\ref{fig:ResVsTemp}, we show the temperature variation of resistance for all three films near $T_c$. The resistances are normalized with respect to maximum values of the respective sample. The room temperature sheet resistance, $R_{\Box}$, the thicknesses, d, and the Ioffe-Regel parameter, $k_F \ell$, are summarized in Table-\ref{tab1}. The temperatures $T_{c1}$s, where resistances go to zero, are also listed in Table-\ref{tab1}. $k_F \ell$ are determined from the electrical resistivity, $\rho_{xx}$,  and Hall coefficient, $R_H$, at room temperature  (see Appendix for thickness and  $k_F \ell$ determination). The $k_F \ell$ values indicate that our films are in the strongly disordered limit \cite{lee1985disordered}.

\begin{figure}\centerline{\includegraphics[width=7cm,angle=0]{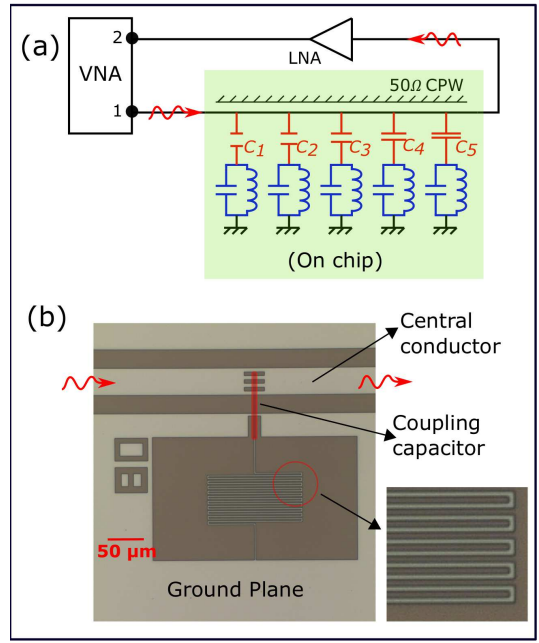}}
\caption {(a) A schematic of our microwave transmission measurement. Five identical microwave resonators are side coupled capacitively to a 50 $\Omega$ matched transmission line with different capacitances. Here, LNA and VNA stand for low noise amplifier and vector network analyzer, respectively. (b) An optical micrograph image of a resonator and a part of the transmission line. The coupling capacitor is shown by a false red-colour. The resonator is meander-shaped with linewidth 2 $\mu m$. A zoomed-in portion of the resonator is shown in the right-bottom of Fig.b. }
\label{fig:Resonator}
\end{figure}

To probe the microwave response of the samples, five identical quasi--lumped-element microwave resonators are fabricated from each film. These resonators are coupled to a 50 $\Omega$ matched transmission line (TL) with varying coupling capacitances, as shown schematically in Fig.\ref{fig:Resonator}a, to probe the microwave response through the transmission, $S_{21}$, measurement. We also show a simplistic schematic diagram of our experimental setup. In Fig.\ref{fig:Resonator}b, we show optical micrograph image of a resonator, coupling capacitor, and part of the TL. For all three samples, the central conductor of the TL is fabricated from 300 nm thick NbN layer. Our lumped-element resonators are meander-shaped with a linewidth of 2 $\mu$m. A zoomed-in portion of the resonator is shown in the right-bottom of Fig.\ref{fig:Resonator}b. The coupling capacitance is provided by sandwiching a $Si_{3}N_{4}$ layer between the central conductor of the TL and a top electrode connecting the resonator, as shown in Fig.\ref{fig:Resonator}b. 

\begin{figure}\centerline{\includegraphics[width=9cm,angle=0]{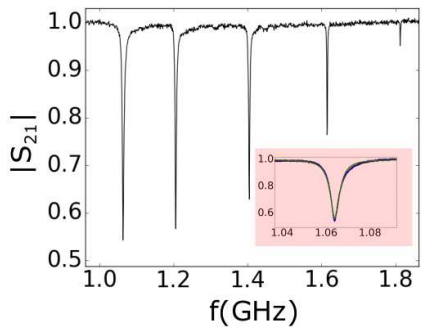}}
\caption {Microwave transmission, $\lvert S_{21} \lvert $, data of S1 at 4.2 K. Five dips correspond to the different coupling capacitances, $C_c$. The inset shows the zoomed-in version of the dip corresponding to the lowest $f_0$. The solid line is a fit with Eq.\ref{eq:S21}
.}
\label{fig:ResonancePeaks}
\end{figure}

In Fig.\ref{fig:ResonancePeaks}, we plot $\lvert S_{21} \lvert $ for S1 at T = 4.2 K. As expected, we observe five fundamental resonances (dips) corresponding to five different coupling capacitors: the lowest $f_0$  corresponds to the highest capacitance and vice-versa. A magnified version of the resonance dip, that corresponds to the lowest $f_0$, is shown in the inset. The $f_0$ is extracted from fitting the resonance curve with the following equation \cite{khalil2012analysis}: 

\begin{eqnarray}
S_{21} = \frac{1+2iQ_{i}\frac{\Delta f_{0}}{f_{0}}}{1+2iQ_{i}\frac{\Delta f_{0}}{f_{0}}+\frac{Q_{i}}{Q_{c}}},
\label{eq:S21}
\end{eqnarray}

here, $Q_i$ and $Q_c$ are internal and coupling quality factor, respectively, and $\Delta f_{0} = f - f_{0}$.

To study the temperature dependence of $f_0$, the first four resonance dips of each sample are traced as a function of temperature. In the limit $L_{k} \gg L_{g}$, where, $L_{k}$ is the kinetic and $L_{g}$ is the geometric inductance of the resonator, $f_0$ can be expressed, in terms magnetic penetration depth $\lambda$, as (see Appendix for detail)

\begin{eqnarray}
\frac{f_{0}(T)}{f_{0}(0)} =  \frac { \lambda(0)}{\lambda(T)},
\label{eq:resFreq1}
\end{eqnarray}

here, $f_{0}(0)$ is the resonance frequency at T = 0.

In the dirty limit --- i.e, $\ell \ll \xi$, where $\ell$ is elastic mean free path $\sim$ 0.5 nm and  $\xi$ is the coherence length $\sim$ 5 nm for our samples as measured from Hall measurement and magnetoresistance data near $T_c$ ---  the BCS theory results \cite{tinkham1996introduction}

\begin{eqnarray}
\frac { \lambda(T)}{\lambda(0)} =  \frac{1}{\sqrt{\delta(T)tanh \left(\beta \delta(T) \frac{T_{c}}{T} \right)}} 
\label{eq:lambda}
\end{eqnarray}
, here, $\delta(T) = \Delta(T)/\Delta(0)$ is normalized energy gap and $\beta  = \Delta(0)/2k_{B}T_{c}$.

Combining Eq.\ref{eq:resFreq1} and Eq.\ref{eq:lambda} yields
 \begin{eqnarray}
f_{0}(T) = f_{0}(0) \sqrt{\delta(T)tanh \left(\beta \delta(T) \frac{T_{c}}{T} \right)}.
\label{eq:resFreq}
\end{eqnarray}

\begin{figure}\centerline{\includegraphics[width=7cm,angle=0]{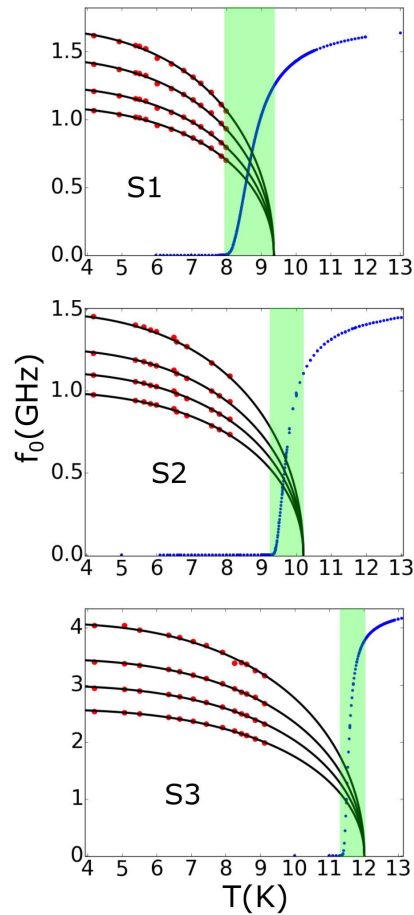}}
\caption {Experimentally obtained resonance frequency (the red dots), $f_0$, as a function of temperature. The solid lines are fits with Eq.\ref{eq:resFreq}. The temperatures, $T_{c}$s, where $f_0$ go to zero, are listed in Table-\ref{tab1}. We also plot the temperature variation of resistance (the blue dots).}
\label{fig:ResistanceFreqTogetherr}
\end{figure}

To verify the agreement between Eq.\ref{eq:resFreq} and our experimental data, we show the temperature dependence of $f_0$ for all three samples in Fig.\ref{fig:ResistanceFreqTogetherr}. In the same figure, we also plot the temperature dependence of resistance near $T_c$, for comparison. The solid lines are fit with Eq. \ref{eq:resFreq}, where, for the temperature dependence of $\delta$, we have simulated it using BCS theory. $f_{0}(0)$ and $T_c$ are taken as fit parameters. The $T_{c}$, as extracted from the fit, are listed in  Table-\ref{tab1}. We see that $T_{c}$, for all three samples, differ significantly from $T_{c1}$ where resistance is zero. In other words, $f_0$ and resistance vanish at different temperatures, $T_{c}$ and $T_{c1}$, respectively. Moreover, the temperature region $T_{c} - T_{c1}$, shown by shaded area, systematically increases with disorder(see Table-\ref{tab1}).

Here, we would like to mention two important points. Firstly, to fit with Eq.\ref{eq:resFreq}, if we set $T_{c} = T_{c1}$ and take $f_{0}(0)$ as the sole fitting parameter, we see that no meaningful fit is possible, especially for S1 and S2. Secondly, for $\beta (= \Delta(0)/2k_{B}T_{c}$), we have chosen $\beta = 1.1$ for all three samples. Though, we have not experimentally determined the $\Delta(0)$ for the reported samples, tunnelling experiments on NbN/MgO/NbN tunnel junction resulted in $\beta$ = 1.1  for similar kinds of films we deposited \cite{grimm2015josephson,grimm2017self}. This number is consistent with Reference \cite{mondal2011PRL, chand2012thesis} for sputtered deposited NbN films with similar $T_c$s. Further, STS measurement on a film identical to S3 resulted $\beta$ = 1.1 (see Appendix). Here, we would like to point out that for disordered superconductors like ours, a spatially inhomogeneous superconducting state may be developed \cite{sacepe2008prl, kamlapure2013emergence,coumou2013PRB}. In that case, $\beta$ refers to a spatially averaged parameter.

\section{Discussion}
To understand the observed behaviour of Fig.\ref{fig:ResistanceFreqTogetherr}, first, we concentrate on the least disordered sample, S3. We note that $f_0$, in the BCS framework, mimics energy gap parameter $\delta$ via Eq.\ref{eq:resFreq}. Thus, at the first instance, it appears that there exists a region, shown by the shaded area, where the global superconductivity is destroyed, as evident from the existance of a finite resistance, but exists finite 'average gap'--- the pseudogap. To verify this, we perform STS measurement on a film identical to S3, deposited under the identical condition that resulted in exactly the same $T_c$ as S3. We didn't observe any pseudogap or spatial inhomogeneity in this film (see Appendix). Here, we would like to point out that several groups have probed disordered NbN films by STS studies. No pseudogap was observed, with best of our knowledge, for films with  $T_c < 6.5 K $ \cite{mondal2011PRL,aberkane2014nano, noat2013unconventional}. Therefore, it is unlikely that S3, with $T_c \sim 11.5 K $, possesses any pseudogap-like features. 

To explore the other possible explanation for the observed behaviour of Fig.\ref{fig:ResistanceFreqTogetherr}, we note that $f_0$ is related to $\delta$ via $\lambda$. Thus, if $\lambda$ deviates from the BCS temperature variation (Eq.\ref{eq:lambda}) and 1/$\lambda$ does have a faster downturn near $T_c$, the temperature variation of $f_0$ for S3 is explicable. Such an observation, indeed, was made for both 2D and 3D disordered NbN films in direct $\lambda$ measurement \cite{kamlapure2010APL, mondal2011PRL, ganguly2015slowing} and was associated, for 2d films, with a BKT type transition that predicts a jump in superfluid density, $n_s (\propto 1/\lambda^{2} \propto f_{0}^{2}) $, at a temperature $T_{BKT}$ before $T_c$. Here, we should point out that in the above references the authors also observed additional smearing, in the temperature dependence of $1/\lambda^{2}$, which they attributed to the vortex core energy \cite{benfatto2007kosterlitz, benfatto2008doping}.

For S1 and S2, since they are thinner than S3 and more disordered, a faster downturn of $n_s$ near $T_c$ are more probable \cite{kamlapure2010APL, mondal2011PRL, ganguly2015slowing, benfatto2007kosterlitz, benfatto2008doping}. However, for S1, we note that $f_0$ is still finite at $T_{c1}$. Thus, in this case, a mere downturn of $n_s$ near $T_c$ is not sufficient to explain our experimental data. Since we didn't perform any STS experiment on films identical to S1 and S2, we cannot completely rule out the possibility of a pseudogap-like features in S1 and S2. Thus, for S1 and S2, particularly for S1, it is possible that both a pseudogap-like feature as well as a downturn of $n_s$ are responsible for the observed behaviour of Fig.\ref{fig:ResistanceFreqTogetherr}. However, a further experimental probe near $T_c$, by STS and direct $\lambda$ measurements, is required to have a final conclusion on that.  

 Finally, we would like to mention that Driessen et al.\cite{driessen2012PRL} reported temperature dependent $f_{0}$ in disordered NbTiN and TiN thin films. With increasing disorder, like us, they also observed a systematic departure from BCS behaviour. However, they observed that $f_{0}$ extrapolates to zero in BCS framework at a temperature lower than where resistance goes to zero (i.e, $T_{c1} > T_{c} $), which is clearly in contradiction to our observation. The main reason behind this apparent contradiction is, in their study, Driessen et al. restricted themselves to very low temperatures, typically below 0.35 $T_{c}$. In that temperature range, using a 3 dimensional model of pair-breaking mechanism, they argued that a modification in the quasiparticle density of state, due to disorder-induced electronic inhomogeneity, is responsible for the observed departure from BCS behaviour. In our case, on the contrary, we explore the temperature range above 0.4 $T_{c}$. Moreover, our films are in the 2-dimensional limit. In our case, therefore, it is possible that BKT like transition dominates over the effect due to the possible modification of the quasiparticle density of state.

\section{Summary and Conclusion}

In summary, we have probed disordered s wave superconductor, three NbN thin films, with Ioffe-Regel parameters, $k_{F} \ell < 4.1$, by electrical transport and microwave signal. Microwave resonators are fabricated and temperature dependence of their resonance frequency, $f_0$, is traced in the superconducting state at temperatures above $0.4T_{c}$. For our measured films, $f_0$ extrapolates to zero, in the BCS framework, at a temperature $T_c$ which is always higher than $T_{c1}$ where resistivity goes to zero. To explain our experimental observation, we discuss the possibility of a pseudogap-like feature as well as the possibility of a faster downturn, than that is expected from the BCS theory, of the temperature dependence of $n_s$ near $T_c$. We, however, advocate for further experiments, near $T_c$, to have a final conclusion.   

\section{Acknowledgements}

We acknowledge financial support from the French National Research Agency, Grant No. ANR-14-CE26-0007–WASI and
Grant No. ANR-16-CE30-0019-ELODIS2, from the Grenoble Fondation Nanosciences, Grant JoQOLaT; and from the European
Research Council under the European Union’s Seventh Framework Programme (FP7/2007-2013), ERC Grant Agreement
No. 278203–WiQOJo.

\section{Appendix}

\subsection {Thin film characterization}

We have deposited a series of NbN thin-films of thicknesses in the range $300 < d <  10 $ nm. The electrical resistivity and Hall measurements are performed on all the films. For microwave transmission measurements, the resonators are fabricated on the three thinnest films where $L_{k} \gg  L_{g} $  and $f_0$ can be expressed in the normalized form of Eq.\ref{eq:resFreq}.

The thicknesses, $d$, of the thicker films,  $d \geq$ 10 nm, are first estimated from optimized deposition rate and then confirmed by the cross-sectional image of high resolution scanning electron micrograph. For some films, the thicknesses are reconfirmed by profilometry measurement. In this way, we could determine film thickness within 10 to 15 \% accuracy.

The electron density, $n$, for the thicker films are determined from the Hall coefficient, $R_{H}$, at room temperature, using the formula: $R_{H} = 1/n e$, here, $e$  is the charge of the electron. The $k_{F}\ell$s are determined from the following formula \cite{hazra2018superconducting}, assuming free electron model: 

\setcounter{equation}{0}
\renewcommand{\theequation}{A\arabic{equation}}
\begin{eqnarray}
k_{F}\ell =\frac{\hbar (3\pi^{2})^{2/3}}{n^{1/3}\rho e^{2}}
\label{eq:resistivity}.
\end{eqnarray}
Here, $\rho$ is the longitudinal electrical resistivity and $\hbar$ is the reduced Planck constant.

To estimate the thicknesses of the thinner films, S1 and S2, we follow two methods:

(1) We note that, for identical deposition condition, $n$ does not change significantly with $d$. This observation is consistent with reference \cite{chand2012thesis}. For our films with known $d$, from Hall measurements at room temperature, we estimate $n \sim  2.0 \times 10 ^ {29} / m ^{3}$. This, in combination with the slope of the Hall resistance $( = 1/n e d )$, is used to estimate the thicknesses of S1 and S2.

(2) Our films with known thicknesses follow the scaling law, recently proposed by Yachin et al. \cite{ivry2014universal}:

\setcounter{figure}{0}
\renewcommand{\thefigure}{A\arabic{figure}}
\begin{figure}\centerline{\includegraphics[width=9cm,angle=0]{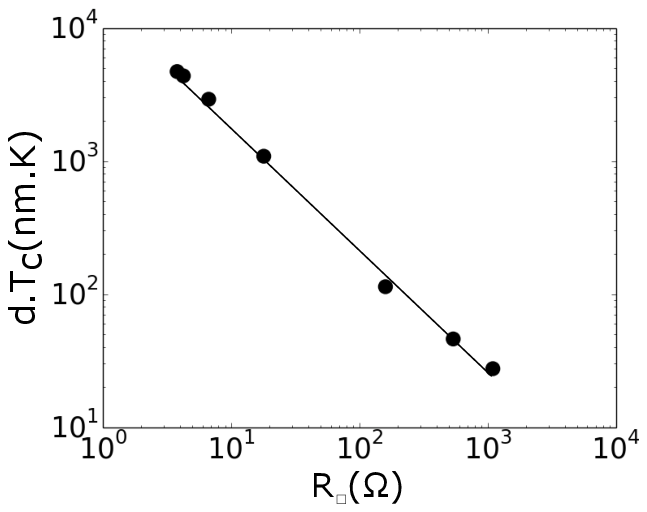}}
	\caption {Variation of $d.T_{c}$ with square resistance $R_{\Box}$ in log scale. The solid line is a straight line fit. }
	\label{fig:TcScalling}
\end{figure}

\begin{eqnarray}
 d.T_{c} = A. R_{\Box}^{-B}
\label{eq:scalling}
\end{eqnarray}

Here, $A$ and $B$ are the fit parameters; $B$ is an exponent close to 1.0. In Fig.\ref{fig:TcScalling}, we show the Variation of $d.T_{c}$ with square resistance, $R_{\Box}$, in log scale. From straight line fit, we extract $A =  1.88 \times 10 ^ {-5}$  and $B$ = 0.99. We assume that S1 and S2 also follow this scaling law with the same values of $A$ and $B$. Knowing $R_{\Box}$, then, we extract thicknesses of S1 and S2, from the straight line fit of Fig.\ref{fig:TcScalling}, in a self-consistent way.

Method 1 and 2 provide consistent results.

Finally, we note that $R_{\Box}$ of S1 and S2 are very similar to d.c. magnetron sputtered deposited films on MgO substrate in Ref.\cite{mondal2011role} for similar thicknesses that resulted similar $T_c$s. 


\subsection{ $f_0$ as a function of temperature}

We consider a resonator, side-coupled with a TL with coupling capacitance $C_c$, like the one shown in Fig.\ref{fig:Resonator}b. It is well known that the resonance frequency ($f_0$) of such a system varies as: $f_0 \propto 1/\sqrt{L_{r}}$ \cite{pozar2009microwave}, where, $L_r$ is the  inductance of the resonator. $L_r$ has both kinetic, $L_k$, and geometric, $L_g$, component: $L_{r} = L_{k} + L_{g}$. Both from the measured $f_0$  as well as by numerical simulation using python and sonnet software, we confirm that for the three samples we have reported, $L_k$ completely overwhelms $L_{g}$. $L_k$ is proportional to surface inductance $L_s$ which is given by $L_{s} = \mu_{0} \lambda^{2}/ d$ for $d \ll \lambda$ \cite{henkels1977penetration,barone1982physics}, here, $\mu_{0}$ is free space permeability. The assumption, $d \ll \lambda$, is justified as  $ d \sim $ 10 nm whereas $\lambda \sim $ 350 nm. This yields Eq.\ref{eq:resFreq1}.

\subsection{ STM measurement}
\begin{figure}\centerline{\includegraphics[width=9cm,angle=0]{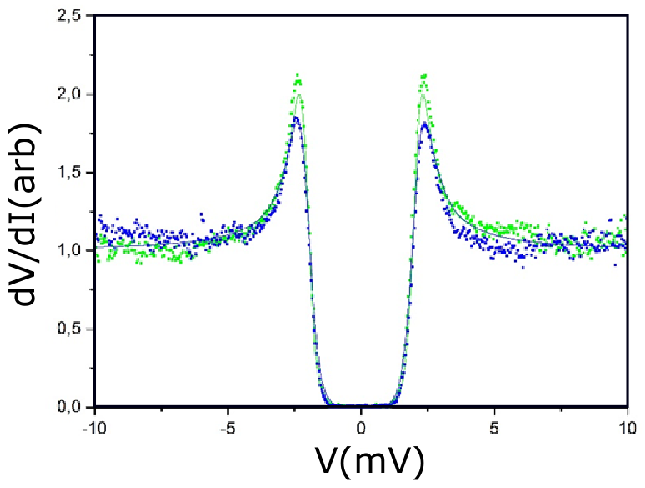}}
	\caption {Scanning tunnelling spectroscopy data at 4.2 K of a film, identical to S3, at the interior of the grain (blue) and grain boundary (green). No difference in $\Delta$ is observed.}
	\label{fig:STM}
\end{figure}

STM measurements are performed on a film of thickness $\sim$ 10 nm, deposited under the identical condition of S1, S2 and S3. STS measurements at 4.2 K on the interior of the grains and grain boundaries reveal that the superconducting gap has non-noticeable variation between grain interiors and grain boundaries, confirming that there is no spatial inhomogeneity due to structural granularity. Also, no distribution of the energy gap parameter is observed. The STS spectra at higher temperatures reveal that there is no pseudogap for this film. In Fig.\ref{fig:STM}, we show two representative spectra taken at 4.2 K.

\bibliography{Bibliography}

\end{document}